\documentclass{appolb}
\usepackage{epsfig}
\begin{document}
% \eqsec  % uncomment this line to get equations numbered by (sec.num)
\title{Properties of localized protons in neutron star matter for realistic nuclear models
\thanks{Presented at the XXIX Mazurian Lakes Conference on Physics, Piaski, Poland,
August 30 - Septenber 6, 2005}
% you can use '\\' to break lines
}
\author{{A. Szmagli\'nski$^a$, W. W\'ojcik$^a$ and M. Kutschera$^{b,c}$}
\address{~$^a$Institute of Physics, Cracow University of Technology,
ul. Podchor\c{a}\.zych 1, 30-084 Krak\'ow, Poland}
\address{~$^b$H. Niewodnicza\'nski Institute of Nuclear Physics, Polish Academy of Sciences,
ul. Radzikowskiego 152, 31-342 Krak\'ow, Poland}
\address{~$^c$M. Smoluchowski Institute of Physics, Jagellonian University,
ul. Reymonta 4, 30-059 Krak\'ow, Poland}
}
\maketitle

\begin{abstract}
We study the localization of protons in the core of neutron stars for ten realistic
nuclear models that share a common behaviour of nuclear symmetry energy
which saturates and eventually decreases at high densities.
This results in the low proton fraction of beta-stable neutron star matter.
Protons form a small admixture in the neutron star core,
which is localized at sufficiently high densities.
For every model we calculate the density $n_{loc}$ above which the localization effect is present.
Our results indicate that localization occurs at densities above $0.5-1.0\,fm^{-3}$.
The phase with localized protons occupies a spherical shell or a core region
inside neutron stars which contains significant fraction of all nucleons.
Proton localization is of great importance for astrophysical properties of neutron stars
as it strongly affects transport coefficients of neutron star matter
and can produce spontaneous magnetization in neutron stars.
\end{abstract}
\PACS{21.65.+f,97.60.Jd}

Observational parameters of neutron stars shed light on some properties of nucleon 
interactions.
These interactions determine the structure and properties of neutron stars.
The transport and magnetic properties of neutron stars depend strongly on the 
structure of the so called liquid core.
Particularly important is the fraction and structure of the proton component.

Nuclear symmetry energy plays a crucial role for composition and other important 
properties of dense matter in neutron star. The interaction part of the nuclear 
symmetry energy $E_S\left( n\right) $
determines the composition of dense nuclear matter \cite{KuPLB}.
Vanishing of the nuclear symmetry energy implies proton-neutron separation instability
in dense nuclear matter \cite{KuPLB}.
Negative values of the symmetry energy result in disappearance of protons at 
high densities. Realistic nuclear matter calculations \cite{WiFiFaPR,Pan/Gar PL B38}
show that the symmetry energy becomes negative at high densities,
and in consequence this leads to the vanishing of protons.

There are two mechanisms of separation of protons and neutrons in 
neutron star matter: a~bulk separation of protons and neutrons and localization of 
individual protons in 
neutron matter. We consider in this paper the latter possibility.
A bulk separation means
that pure neutron matter coexists with nucleon matter containing some proton 
fraction $x_c$.

Protons which form admixture tend to be localized in potential wells
corresponding to neutron matter inhomogeneities created by the protons in the neutron medium.
To compare the energy of a~normal phase of uniform density and a phase with localized protons
we apply the Wigner-Seitz approximation and divide the system into cells,
each of them enclosing a single localized proton \cite{KuWoPL,KuWoPR}.
The neutron background is treated in the Thomas-Fermi approximation
and the localized proton is described by the Gaussian wave function.
The neutron density profile is obtained by solving the appropriate variational equation \cite{Kut/Sta/Szm/Woj APP B33}.
We calculate the difference of the energy per proton in the two phases and look for a 
critical density for localization at which the localized phase becomes the ground state.

The localization effect is a result of the interaction of protons
with small density oscillations of the neutron background \cite{KuWoPR}.
The protons behave as localized polarons which form a periodic lattice at high 
densities \cite{KuWoNP}.The aim of this paper is to study the proton localization for 
a number of realistic nuclear models with selfconsisted variational method.

The amount of protons present in the neutron star matter, which is charge neutral and 
$\beta$-stable,
is crucial for the cooling rate of neutron stars and also plays an important role
for magnetic and transport properties of neutron star matter.
Nuclear models do not uniquely predict the proton fraction of the neutron star matter at high densities.
This controversy is discussed in details in Ref. \cite{KuPLB,KuZP} where the discrepancy of the proton
fraction in various models is shown to reflect the uncertainty of the nuclear symmetry energy at high densities.
In this paper we consider a class of nuclear interaction models for which the proton fraction
is of the order of a few (for five of them of above ten) percent and decreases at high densities.
For the calculations we have chosen ten realistic nuclear interaction models.
These are interactions derived by Myers and Swiatecki \cite{MySwZSP} (MS),
the Skyrme model with parameters (SI', SII', SIII', SL, Ska, SKM, SGII, RATP, T6)
from Ref. \cite{LaARNPS,Ch/Bo/Ha/Me/Sch NP A627},
the Friedman and Pandharipande interactions \cite{FrPaNP} (as parametrized by Ravenhall in Ref.\cite{LaARNPS}) (FPR),
three models, UV14+TNI, AV14+UVII and UV14+UVII, from Ref.\cite{WiFiFaPR} by Wiringa et al.
and four modern APR models \cite{Akm/Pan/Rav PR C58},
which provides a fit to the nucleon-nucleon scattering data in the Nijmegen data base.
Fig.1 shows the proton fraction of beta-stable and charge neutral matter containing neutrons,
protons, electrons and muons for all the interactions.

\begin{figure}[hbt]
	\begin{center}
		\includegraphics[height=6.5cm]{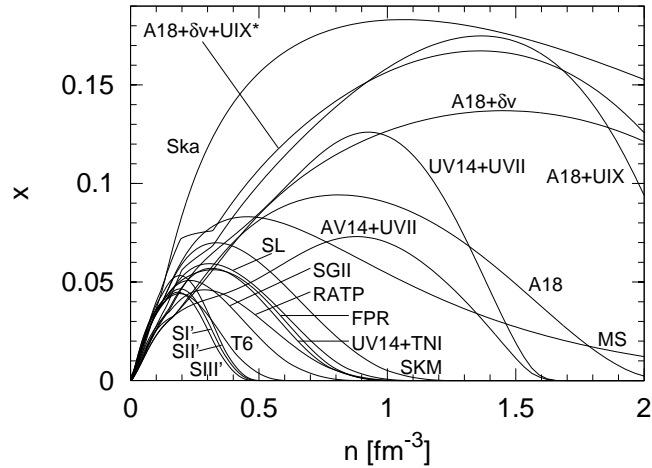}
	\end{center}
	\caption{Proton fraction of neutron star matter for indicated interaction models.}
	\label{x_n}
\end{figure}

In calculactions we apply selfconsisted method described in our paper \cite{Kut/Sta/Szm/Woj APP B33}.
The threshold densities (in $fm^{-3}$) as well as rms proton radii (in $fm$),
when the localization effect becomes energetically favorable are listed in Table 1.
We also present the minimum masses of neutron stars $M_{\min}$,
when central density is equal to the threshold density of localization of protons.
For every model we calculated also the maximum mass of neutron stars $M_{\max}$ (in solar units).
The localization effect is present above the threshold barion number density $n_{loc}$.
For some models protons vanish from the system above some density $n_v$,
so the localization effect is present in the spherical shape region of neutron stars in the density range
$\left\langle n_{loc};n_v\right\rangle $ Fig.2.
For Ska parametrization of the Skyrme model and the A18+$\delta$v model the threshold density
for proton localization  is also the central density of predicted maximum mass configuration of neutron star,
so there is no localization of proton inside the observed neutron stars in these two cases.

Localization of protons influences the global properties of neutron star, so it is of importance 
to calculate the fraction of nucleons inside
the region with localized protons in the neutron star. We show this fraction in Fig.3.

Results of our calculations for nuclear interactions we use indicate that the proton impurity
in neutron star matter becomes localized at densities above $0.5-1.0 fm^{-3}$.
This has important consequences for neutron stars as densities in this range correspond to 
the inner core of neutron stars with masses exceeding one solar mass, $M> 1M_{\odot}$.

\vspace{.5cm}
{\footnotesize {\bf TABLE 1}\\
For considered models we present the threshold densities for localization
and corresponding rms radii of proton wave function calculated using selfconsistent variational method,
the minimum mass of neutron star $M_{\min}$ (in solar units) which has central density equal
to the threshold density of localization and the predicted maximum mass od neutron star.}
\nopagebreak[4]
\vspace{-.5cm}
\nopagebreak[4]

\begin{center}
\begin{tabular}
[c]{||l||c|c|c|c||}\hline\hline
Potential & $n_{loc}$ & $R_{P}^{loc}$ & $M_{\min}$ & $M_{\max}$\\\hline\hline
MS			& 1.030 & 0.905 & 2.019 & 2.038\\\hline
SI'			& 0.351 & 1.519 & 0.876 & 2.242\\\hline
SII'			& 0.361 & 1.688 & 0.829 & 1.966\\\hline
SIII'			& 0.337 & 1.552 & 0.898 & 2.266\\\hline
SL			& 0.964 & 1.384 & 1.480 & 1.631\\\hline
Ska			& 1.016 & 0.804 & 2.241 & 2.241\\\hline
SKM			& 0.979 & 1.330 & 1.570 & 1.677\\\hline
SGII			& 0.899 & 1.367 & 1.432 & 1.661\\\hline
RATP			& 0.709 & 1.397 & 1.225 & 1.726\\\hline
T6			& 0.530 & 1.985 & 0.600 & 1.429\\\hline
FPR			& 0.721 & 1.262 & 1.435 & 1.800\\\hline
UV14+TNI		& 0.731 & 1.209 & 1.489 & 1.827\\\hline
AV14+UVII		& 0.789 & 0.971 & 1.793 & 2.123\\\hline
UV14+UVII		& 0.766 & 0.913 & 1.986 & 2.207\\\hline
A18			& 1.493 & 1.136 & 1.647 & 1.673\\\hline
A18+$\delta$v	& 1.627 & 0.915 & 1.805 & 1.805\\\hline
A18+UIX		& 0.645 & 0.911 & 2.095 & 2.386\\\hline
A18+$\delta$v+UIX*	& 0.819 & 0.878 & 2.052 & 2.213\\\hline\hline
\end{tabular}
\end{center}

\begin{figure}[hbt]
	\begin{center}
		\includegraphics[height=6.5cm]{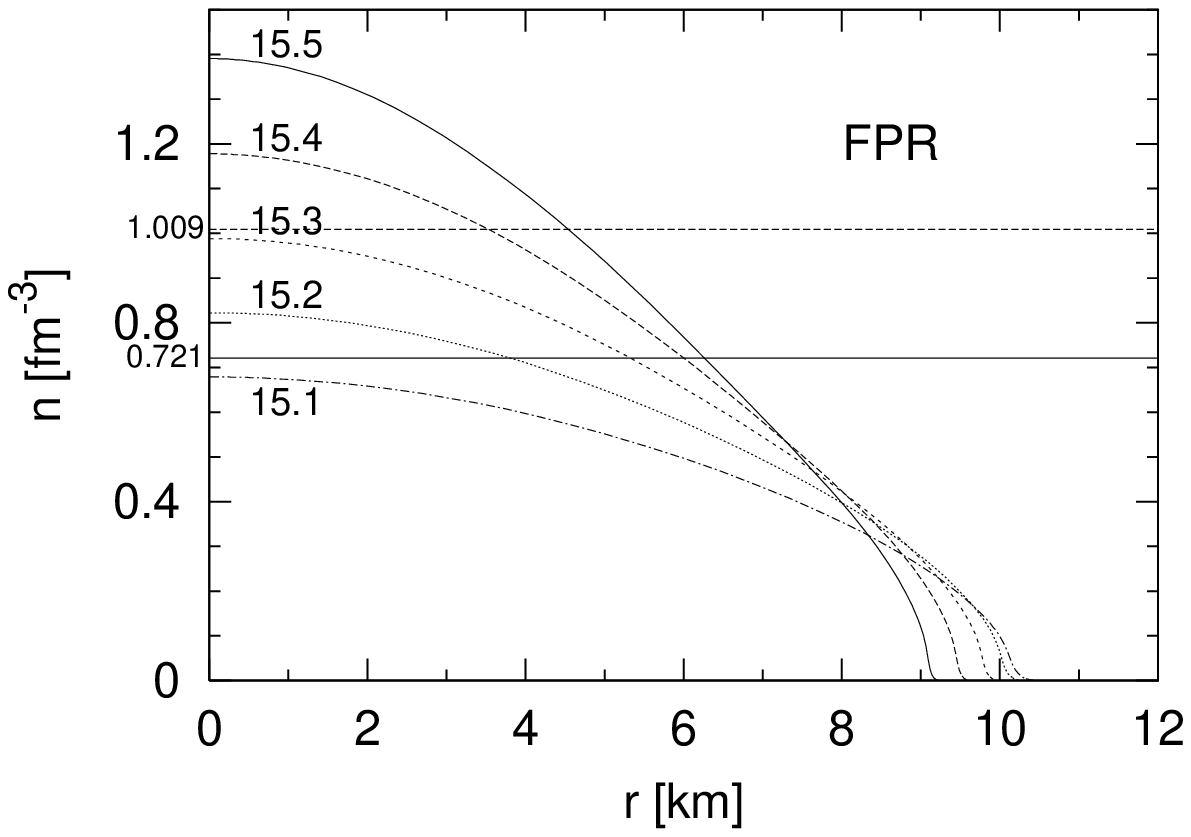}
	\end{center}
	\caption{The spherical shell  of neutron star,
where the localization effect occurs. The region corresponds to density above the threshold
and below the density for which the protons vanish.
Neutron star density profiles are labelled by logarithm of the central density.}
	\label{area}
\end{figure}

\begin{figure}[hbt]
	\begin{center}
		\includegraphics[height=6.5cm]{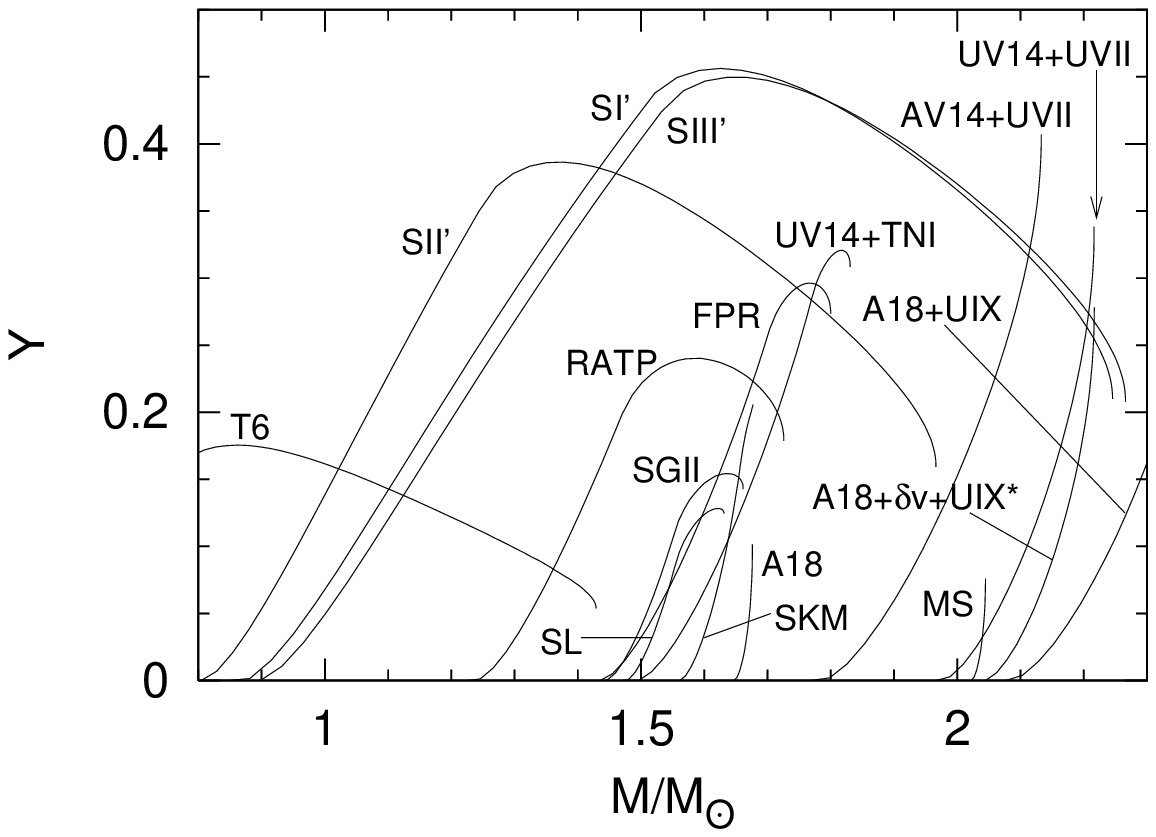}
	\end{center}
	\caption{The fraction of nucleons in the region with localized protons
of the neutron star versus neutron star mass.}
	\label{fraction}
\end{figure}

According to the Ref. \cite{Lat/Pra Sc304} significant number of measured neutron star masses are inside the range
$\left\langle M_{\min};M_{\max}\right\rangle $.
It indicates that localization of protons is an universal state of dense nuclear matter in neutron stars.

Strongly asymmetric nuclear matter is unstable with respect to proton localization \cite{KuWoAPPB21,KuWoPL}.
A uniform proton distribution and a periodic proton arrangement result in very different properties \cite{KuPLB}.
The presence of the localized protons inside neutron star cores would have profound astrophysical consequences.
In particular, the cooling proceeds in a quite different way.
Recent analysis \cite {BaHaAA} shows that the presence of such localized proton phase
results in more satisfactory fits of temperatures of observed neutron stars. 
Also, spin ordering of localized protons could strongly affect magnetic properties of the system
\cite{KuWoAPPB21,KuWoAPPB27}.
The spin ordered phase can contribute significantly to the observed magnetic moments of neutron stars
\cite{KuWoAPPB23,KuMNRAS}.

This research is partially supported by the Polish State Committee for Scientific Research, grants 2P03B 110 24 and
PBZ-KBN-054/P03/2001.


\begin{thebibliography}{99}


\bibitem{KuPLB} M. Kutschera, \ Phys. Lett. {\bf B340}, 1 (1994).
%
\bibitem{WiFiFaPR} R. W. Wiringa, V. Fiks and A. Fabrocini, \ Phys. Rev. {\bf C38}, 1010 (1988).
%
\bibitem {Pan/Gar PL B38}V. R. Pandharipande and V. K. Garde, \ Phys. Lett. {\bf B38}, 608 (1972).
%
\bibitem{KuWoPL} M. Kutschera, W. W\'ojcik, \ Phys. Lett. {\bf B223}, 11 (1989).
%
\bibitem{KuWoPR} M. Kutschera, W. W\'ojcik, \ Phys. Rev. {\bf C47}, 1077 (1993).
%
\bibitem{Kut/Sta/Szm/Woj APP B33}
M. Kutschera, S. Stachniewicz, A. Szmagli\'nski, W. W\'ojcik, \ Acta Phys. Pol. {\bf B33}, 743 (2002).
%
\bibitem{KuWoNP} M. Kutschera and W. W\'ojcik, Nucl. Phys. {\bf A581}, 706 (1995).
%
\bibitem{KuZP} M. Kutschera, \ Z. Phys. {\bf A348}, 263 (1994).
%
\bibitem{MySwZSP} W. D. Myers and W. J. Swiatecki, \ Acta Phys. Pol. {\bf B26}, 111 (1995).
%
\bibitem{LaARNPS} J. M. Lattimer, \ Ann. Rev. Nucl. Part. Sci. {\bf 31}, 337 (1981).
%
\bibitem{Ch/Bo/Ha/Me/Sch NP A627}
E. Chabanat, P. Bonche, P. Haensel, J. Meyer, R. Schaeffer, Nucl. Phys. A {\bf 627}, 710-746 (1997).
%
\bibitem{FrPaNP} B. Friedman and V. R. Pandharipande, \ Nucl. Phys. {\bf A361}, 502 (1981).
%
\bibitem{Akm/Pan/Rav PR C58}
A. Akmal, V.R. Pandharipande and D. G. Ravenhall, \ Phys. Rev. {\bf C58}, 1804 (1998).
%
\bibitem{Lat/Pra Sc304} J. M. Lattimer and M. Prakash, \ Science {\bf 304}, 536 (2004).
%
\bibitem{KuWoAPPB21} M. Kutschera and W. W\'ojcik, \ Acta Phys. Pol. {\bf B21}, 823 (1990).
%
\bibitem{BaHaAA} D. A. Baiko and P. Haensel, \ Astron. Astrophys. {\bf 356}, 171 (2000).
%
\bibitem{KuWoAPPB27} M. Kutschera and W. W\'ojcik, \ Acta Phys. Pol. {\bf B27}, 2227 (1996).
%
\bibitem{KuWoAPPB23} M. Kutschera, W. W\'ojcik, \ Acta Phys. Pol. B {\bf B23}, 947 (1992).
%
\bibitem{KuMNRAS} M. Kutschera, \ MNRAS, {\bf 307}, 784 (1999).

\end{thebibliography}
\end{document}